\documentclass[twocolumn]{aastex62}

\begin{document}

\title{A Candidate Dual QSO at Cosmic Noon}

\author[0000-0003-0489-3750]{Eilat Glikman}
\affiliation{Department of Physics, Middlebury College, Middlebury, VT 05753, USA}

\author[0000-0002-5116-6217]{Rachel Langgin}
\affiliation{Department of Astrophysics, Haverford College, Haverford, PA 19041, USA}
\affiliation{Bryn Mawr College, Bryn Mawr, PA 19010, USA}

\author[0000-0001-7690-3976]{Makoto A. Johnstone}
\affiliation{Department of Physics, Middlebury College, Middlebury, VT 05753, USA}

\author[0000-0001-9163-0064]{Ilsang Yoon}
\affiliation{National Radio Astronomy Observatory, 520 Edgemont Road, Charlottesville, VA 22903, USA}

\author[0000-0001-8627-4907]{Julia M. Comerford}
\affiliation{Department of Astrophysical and Planetary Sciences, University of Colorado, Boulder, CO 80309, USA}

\author[0000-0001-5882-3323]{Brooke D. Simmons}
\affiliation{Department of Physics, Lancaster University, Bailrigg, Lancaster LA1 4YB, UK}

\author[0000-0002-8999-9636]{Hannah Stacey}
\affiliation{Max Planck Institute for Astrophysics, Karl-Schwarzschild Str 1, D-85748 Garching bei München, Germany}

\author[0000-0002-3032-1783]{Mark Lacy}
\affiliation{National Radio Astronomy Observatory, 520 Edgemont Road, Charlottesville, VA 22903, USA}

\author[0000-0002-7893-1054]{John M.~O'Meara}
\affiliation{Department of Physics, Saint Michael's College, One Winooski Park, Colchester, VT, 05439, USA}
\affiliation{W.M. Keck Observatory 65-1120 Mamalahoa Highway, Kamuela, HI 96743, USA}

\begin{abstract}

We report the discovery of a candidate dual QSO at z=1.889, a redshift that is in the era known as ``cosmic noon'' where most of the Universe's black hole and stellar mass growth occurred. The source was identified in {\em Hubble} Space Telescope WFC3/IR images of a dust-reddened QSO that showed two closely-separated point sources at a projected distance of 0\farcs26, or $2.2$ kpc. This red QSO was targeted for imaging to explore whether red QSOs are hosted by merging galaxies. We subsequently obtained a spatially-resolved STIS spectrum of the system, covering the visible spectral range, and verifying the presence of two distinct QSO components. 
We also obtained high-resolution radio continuum observations with the VLBA at 1.4~GHz (21-cm L band) and found two sources coincident with the optical positions. 
The sources have similar black hole masses, bolometric luminosities, and radio loudness parameters. 
However, their colors and reddenings differ significantly. The redder QSO has a higher Eddington ratio, consistent with previous findings.
We consider the possibility of gravitational lensing and and find that it would require extreme and unlikely conditions. If confirmed as a bona-fide dual QSO, this system would link dust-reddening to galaxy and supermassive black hole mergers, opening up a new population in which to search for samples of dual AGN.
\end{abstract}


\section{Introduction} \label{sec:intro}

The next generation gravitational wave experiment, LISA, will detect the signal from the coalescence of supermassive black holes (SMBHs) in the $10^{5} - 10^{7} M_\odot$ range.
Since every large galaxy hosts a nuclear SMBH, understanding the black hole merger process into the supermassive regime is essential for a full picture of galaxy evolution. 
Galaxy mergers have also been invoked to explain the many scaling relations seen between galaxies and their nuclear supermassive black holes (SMBHs) suggesting a co-evolution between the two systems \citep{Magorrian98,Gebhardt00,Marconi03}. 
In addition, gas-rich (i.e., ``wet'') mergers are understood to trigger the most luminous QSOs through the funneling of gas and dust into to the nucleus fueling accretion onto the SMBHs, which are also being brought together by the merger \citep{Sanders88}. 
It is expected, therefore, that at some point during this process both SMBHs will be simultaneously active and therefore discoverable as a pair of active galactic nuclei (AGNs).

While theoretical investigations into the physics of SMBH binaries (e.g., mass ratio, coalescence time scale, AGN activity) have been making steady progress, observational constraints are still lacking due to the small numbers of confirmed dual AGNs. These simulations do find that late-stage major mergers are the most likely to produce dual AGNs \citep[i.e., separations $\le$ 10 kpc;][]{vanWassenhove12,Blecha13,Steinborn16} suggesting that those are the best systems in which to search.

Dust-reddened (or, red) QSOs, represent a short-lived phase of QSO evolution driven by the “wet” merger scenario described above. During such a merger, much of the black hole growth occurs in a heavily enshrouded environment
followed by a relatively brief transitional phase in which the obscuring dust is cleared by outflows and radiation-driven winds and is seen as a moderately reddened, Type 1, luminous QSO. After feedback processes clear the dust, the canonical blue QSO shines through and dominates \citep{Sanders88,Hopkins05,Hopkins08}. 
Objects in the transitional phase, i.e, moderately obscured, red QSOs, are farther along the merger timeline, and are thus ideal systems for finding dual AGNs.

Samples of red quasars\footnote{In this letter, we adopt the canonical nomenclature that distinguishes quasars, radio-detected luminous AGN whose radio emission is essential to their selection, from QSOs, the overall class of luminous AGN.} have been identified through radio plus near-infrared selection \citep[e.g., the FIRST-2MASS, or F2M, red quasar survey;][]{Glikman04,Glikman07,Glikman12} and, more recently, mid- plus near-infrared selection \citep[e.g., the WISE-2MASS, or W2M, red QSO survey;][]{Glikman18,Glikman22}. 
These red QSO samples span a broad range of redshifts ($0.1 < z < 3$) and reddenings ($0.2 \lesssim E(B - V) \lesssim 1.5$); have very high accretion rates \citep[$L/L_{\rm Edd} > 0.1$;][]{Kim15}, sufficient to blow out the obscuring material \citep{Glikman17}. 
Their spectra often show broad absorption lines (BALs) that are associated with outflows and feedback \citep{Urrutia09}. 
Crucially, {\em Hubble} Space Telescope (HST) imaging at $z \simeq 0.7$ and $z \simeq 2$ reveals that $\gtrsim 80\%$ of F2M red quasars are hosted by merging galaxies \citep{Urrutia08, Glikman15} making them more likely to harbor dual AGNs (or, more luminous, dual QSOs).

In this paper, we present the discovery of a candidate dual QSO in HST imaging of a sample of W2M red QSOs from \citet{Glikman22}. The QSO's redshift of $z\sim1.9$ probes the epoch of peak AGN and star formation in the universe.
Throughout this work we quote magnitudes on the AB system, unless explicitly stated otherwise. 
When analyzing spectra for extinction properties, we first correct them for Galactic extinction, using the \citet{F99} extinction curve.
When computing luminosities and any other cosmology-dependent quantities, we use the $\Lambda$CDM concordance cosmology: $H_0 = 70$ km s$^{-1}$ Mpc$^{-1}$, $\Omega_M = 0.30$, and $\Omega_\Lambda = 0.70$. 

\section{Source Characteristics} \label{sec:discovery}

\begin{figure*}[ht!]
\includegraphics[scale=0.44]{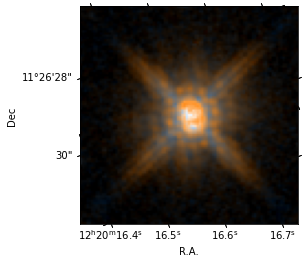}
\includegraphics[scale=0.56]{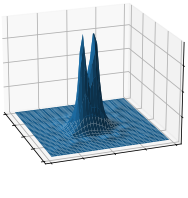}
\includegraphics[scale=0.35]{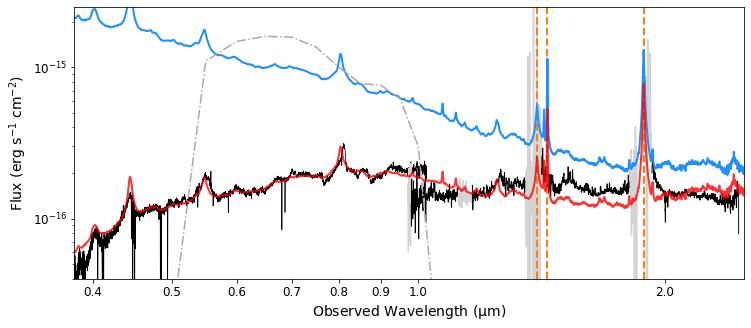}
\caption{{\em Left --} Color combined WFC3/IR image showing the presence of two closely-separated central peaks. The red layer is the F160W image, the green layer is an average of the F160W and F105W images, and the blue layer is the F105W image. 
{\em Middle --}  Surface plot of the image counts in the F160W image where two distinct sources are visible.
{\em Right --} Optical through near-infrared spectrum of W2M1220+1126 (black line). A reddened QSO template, made out of the UV composite QSO template of \citet{Telfer02} combined with the optical-to-near-infrared composite spectrum from \citet{Glikman06}, with $E(B-V) = 0.246$ is overplotted with a red line and an unreddened QSO template is shown in blue. We see that the Balmer lines are shifted into the atmospheric absorption bands. The STIS G750L transmission curve used in this work is shown with a gray dot-dash curve.
\label{fig:discovery}}
\end{figure*}

A Cycle 24 HST program imaged the host galaxies of 11 W2M red QSOs (5 QSOs at $z\sim0.7$ and 6 QSOs $z\sim2$) to compare with the F2M imaging studies of \citet{Urrutia08} and \citet{Glikman15} that focused on those same redshifts (13 and 11 objects, respectively), using ACS and WFC3/IR, respectively (PID 14706, PI Glikman). 
The images were observed with a four-point box dither pattern and were reduced using the {\tt Astrodrizzle} package with a final pixel scale of 0\farcs06.
One source, J122016.9+112627.09\footnote{Hereafter, W2M~J1220}, appeared as two closely separated point sources (left and middle panels of Figure \ref{fig:discovery}) in both the F105W and F160W filters.
The WFC3/IR observations were designed to be identical to those in \citet{Glikman15}, with W2M~J1220 having exposure times of 797 s and 1597 s and reaching 3$\sigma$ surface brightness limits of 23.67 mag arcsec$^{-2}$ and 23.87 mag arcsec$^{-2}$ in F105W and F160W, respectively. 
From the ground, this source appears as a single object with $r=18.13$ in the Sloan Digital Sky Survey (SDSS) and $H=16.01$ in 2MASS. 

This source possesses an optical spectrum in SDSS and was assigned a redshift of $z=1.871$ (see \S \ref{sec:qso} for details on the corrected redshift), shown in the right panel of Figure \ref{fig:discovery}. The spectrum is well-fit by a QSO composite spectrum, constructed by combining the UV template of \citet{Telfer02} with the optical-to-near-infrared template from \citet{Glikman06} and reddened with the SMC dust law of \citet{Gordon98}, by $E(B-V) = 0.246$. We also obtained a near-infrared spectrum with the TripleSpec near-infrared spectrograph \citep{Wilson04} on the 200-inch Hale telescope at the Palomar Observatory, also shown in Figure \ref{fig:discovery}. The Balmer lines are shifted into the atmospheric absorption bands and cannot be studied from the ground. Due to the seeing-limited resolution of $\sim 1\arcsec$, this optical-through-near-infrared spectrum represents the combined light of both components and is therefore not well-fit by a single reddened QSO across the full wavelength range. 

W2M~J1220 is also detected in the FIRST survey with an integrated flux density of $F_{\rm int,20cm} = 2.33$ mJy ($F_{\rm pk,20cm} = 1.50$ mJy/beam), which corresponds to a total radio power of $P_{1.4{\rm GHz}} = 5.7 \times 10^{25}$ W Hz$^{-1}$.
Table \ref{tab:phot} lists the optical through near-infrared photometry for this source.

%% LaTeX deluxetable generator for the AASTeX package.
%% Written by Greg Schwarz (5/1/2001).

%% Table generated: Wed Jul 20 12:38:15 2022

%% The values (usually only l,r and c) in the last part of
%% \begin{deluxetable}{} command tell LaTeX how many columns
%% there are and how to align them.
\begin{deluxetable}{cc}

%% Keep a portrait orientation

%% Over-ride the default font size
%% Use Default (12pt)

%% Use \tablewidth{?pt} to over-ride the default table width.
%% If you are unhappy with the default look at the end of the
%% *.log file to see what the default was set at before adjusting
%% this value.
%% This is the title of the table.
\tablecaption{Photometric properties of F2M J1220} 

%% This command over-rides LaTeX's natural table count
%% and replaces it with this number.  LaTeX will increment 
%% all other tables after this table based on this number
%\tablenum{1}

%% The \tablehead gives provides the column headers.  It
%% is currently set up so that the column labels are on the
%% top line and the units surrounded by ()s are in the 
%% bottom line.  You may add more header information by writing
%% another line between these lines. For each column that requires
%% extra information be sure to include a \colhead{text} command
%% and remember to end any extra lines with \\ and include the 
%% correct number of &s.
\tablehead{\colhead{Band} & \colhead{AB mag}
} 
%% All data must appear between the \startdata and \enddata commands
\startdata
g      &  19.07$\pm$0.01 \\ 
r      &  18.130$\pm$ 0.008 \\ 
i      &  17.452$\pm$0.007 \\ 
z      &  17.12$\pm$0.01 \\ 
J      &  16.44$\pm$0.07 \\ 
H      &  16.01$\pm$0.06 \\ 
K      &  15.67$\pm$0.06 \\ 
\hline
\multicolumn{2}{c}{Northern Component$^a$}\\
\hline
F105W  &  18.020$\pm$0.003 \\ 
F160W  &  16.9739$\pm$0.0002 \\ 
\hline
\multicolumn{2}{c}{Southern Component$^a$}\\
\hline
F105W  &  17.411$\pm$0.002 \\ 
F160W  &  16.8656$\pm$0.0004 \\ 
\enddata

%% Include any \tablenotetext{key}{text}, \tablerefs{ref list},
%% or \tablecomments{text} between the \enddata and 
%% \end{deluxetable} commands
\tablenotetext{a}{The magnitudes reported from the HST images are based on the PSF components determined by Galfit as described in \S \ref{sec:galfit}.}
%% No \tablecomments indicated

%% No \tablerefs indicated
\label{tab:phot} 

\end{deluxetable}

\subsection{Morphological modeling}\label{sec:galfit}

To determine the separation of the two sources and measure their respective magnitudes we modeled the WFC3/IR images in both filters using Galfit \citep{Peng02}. 
We used a point spread function (PSF) constructed by combining a few dozen bright stars in each HST filter from observations that used the same dither pattern following the same procedure as in \citet{Glikman15} using stars whose images were were obtained within 12 months of W2M~1220. The stars used to construct the PSF were chosen to lie in the central region of the WFC3/IR detector to minimize distortion effects. All archival observations were re-reduced using the same {\tt Astrodrizzle} parameters as W2M~1220. 
When fitting, all parameters were allowed to be free in both filter images, which produced the best fits (i.e., smallest $\chi^2_\nu$) and cleanest residuals. 
Attempts at fixing the positions of the F160W and F105W components to each other resulted in poorer fits and yielded residuals with strong negative/positive flux asymmetries. 

We first fit a model consisting of two PSFs and a background sky component.
While both sources are consistent with a point spread function, the residual image showed excess flux in need of additional model components. We added a S\'{e}rsic component to the model resulting in a better fit, verified by an F-test whose probability was consistent with 0, strongly suggesting that we can reject the null hypothesis. 
However, the best-fit S\'{e}rsic component, situated next to the southern PSF, had an effective radius, $R_e$, of 0.06 pixels, which is not physically meaningful. 
Although the addition of this S\'{e}rsic component improved the fit statistic and accounted for flux not captured by the PSFs, it is unclear how much of this added component was accounting for PSF mismatches\footnote{Extensive investigation into the PSF subtraction did not reveal any systematic effects when fit to archival point sources located at the same pixel position as W2M~1220. However, we find that the PSF is not able to capture all the flux from very bright point sources resulting in significant, yet symmetric residuals.}.
Because the added S\'{e}rsic did not model the extended emission seen in the residual image to the east of the two PSFs, we added a second S\'{e}rsic component, which does account for this excess flux and whose inclusion is supported by an F-test with probability consistent with 0. 

\begin{figure*}[ht!]
\begin{center}
\includegraphics[scale=0.5]{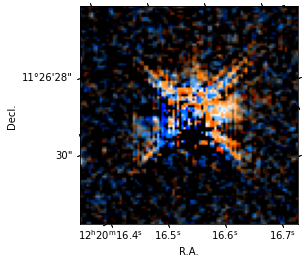}
\includegraphics[scale=0.5]{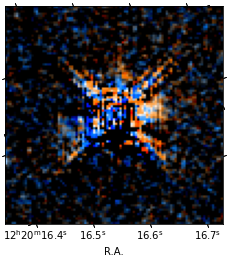}
\includegraphics[scale=0.5]{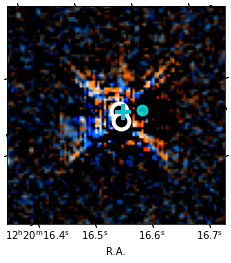}
\caption{Color combined residual images from the best-fit Galfit model, as described in Table \ref{tab:galfit}. 
{\em Left --} HST WFC3/IR image with just the two PSF components subtracted. 
{\em Middle --} Both PSF components and central S\'{e}rsic component subtracted; bright extended emission is seen to the east.
{\em Right --} Full residual with all model components subtracted. In this frame, the best-fit model parameters are marked. White circles are at the PSF positions. The cyan cross is the central S\'{e}rsic parameter located slightly to the east of the southern PSF. The cyan circle is the position of the S\'{e}rsic component that best-fits the extended emission farther to the east.    
\label{fig:galfit}}
\end{center}
\end{figure*}

The best-fit Galfit model therefore is composed of two PSFs and two S\'{e}rsic components. The locations of the PSFs, in both filters, indicate a projected separation of $0.2680\pm0.0003\arcsec$, which corresponds to $\sim 2.2$ kpc at the QSO’s redshift. 
Figure \ref{fig:galfit} shows the residual images from this fit where different model components have been subtracted from the data. We mark in the rightmost panel the positions of the four model components.
Table \ref{tab:galfit} lists the best-fit parameters for this model, noting that the first S\'{e}rsic component may not represent a physically meaningful model.

%% LaTeX deluxetable generator for the AASTeX package.
%% Written by Greg Schwarz (5/1/2001).

%% Table generated: Mon May 15 12:36:15 2023

%% Remove the two lines and the last line if you want
%% want to incorporate this table into another LaTex document.
%\documentclass{aastex}
%\begin{document}

%% The values (usually only l,r and c) in the last part of
%% \begin{deluxetable}{} command tell LaTeX how many columns
%% there are and how to align them.
\begin{deluxetable*}{ccccccccccc}

%% Rotate to a landscape orientation
%\rotate

%% Over-ride the default font size
%% Use Default (12pt)

%% Use \tablewidth{?pt} to over-ride the default table width.
%% If you are unhappy with the default look at the end of the
%% *.log file to see what the default was set at before adjusting
%% this value.

%% This is the title of the table.
\tablecaption{Galfit parameters}

%% This command over-rides LaTeX's natural table count
%% and replaces it with this number.  LaTeX will increment 
%% all other tables after this table based on this number
%%\tablenum{2}

%% The \tablehead gives provides the column headers.  It
%% is currently set up so that the column labels are on the
%% top line and the units surrounded by ()s are in the 
%% bottom line.  You may add more header information by writing
%% another line between these lines. For each column that requries
%% extra information be sure to include a \colhead{text} command
%% and remember to end any extra lines with \\ and include the 
%% correct number of
\tablehead{
\colhead{Component} & \multicolumn{5}{c}{F160W ($\chi^2_\nu = 58.2$)} & \multicolumn{5}{c}{F105W ($\chi^2_\nu = 28.8$)} \\
\colhead{} & \colhead{R.A.\tablenotemark{a}} & \colhead{Decl. \tablenotemark{a}} & \colhead{Mag} & \colhead{$n$} & \colhead{$R_e$} & \colhead{R.A.\tablenotemark{a}} & \colhead{Decl.\tablenotemark{a}} & \colhead{Mag} & \colhead{$n$} & \colhead{$R_e$} \\ 
\colhead{} & \colhead{(J2000)} & \colhead{(J2000)} & \colhead{(mag)} & \colhead{} & \colhead{(kpc)} & \colhead{(J2000)} & \colhead{(J2000)} & \colhead{(mag)} & \colhead{} & \colhead{(kpc)} } 

%% All data must appear between the \startdata and \enddata commands
\startdata
North PSF            & +0.8800  & +0.2805  & 16.97  & \ldots        & \ldots        & +0.8783  & +0.2875  & 18.02  & \ldots       & \ldots    \\ 
South PSF            & +0.8785  & +0.0097  & 16.87  & \ldots        & \ldots        & +0.8781  & +0.0232  & 17.41  & \ldots       & \ldots    \\ 
S\'{e}rsic (central) & +0.8922  & +0.0668  & 20.31  & $1.32\pm6.21$ & $0.03\pm0.23$ & +0.8926  & +0.0471  & 21.00  & 1.76\tablenotemark{b} & 0.02\tablenotemark{b} \\ 
S\'{e}rsic (eastern) & +0.9379  & +0.1756  & 20.89  & 1.94\tablenotemark{b} & 1.5\tablenotemark{b} & +0.9383  & +0.1469  & 20.31  & $1.50\pm0.37$ & $1.01\pm0.04$ \\ 
\enddata
%% Include any \tablenotetext{key}{text}, \tablerefs{ref list},
%% or \tablecomments{text} between the \enddata and 
%% \end{deluxetable} commands
\tablenotetext{a}{The positions reported are shifts in seconds with respect to R.A. 12:20:16 and shifts in arcseconds with respect to Decl. +11:26:28.}
\tablenotetext{b}{These parameters are flagged by Galfit as being outside the range of acceptable values. However, the fit resulted in an acceptable $\chi^2_\nu$ enabling a capture of the
residual flux in the components. We do not report errors for these parameters.}
%% No \tablecomments indicated

%% No \tablerefs indicated
\label{tab:galfit}

\end{deluxetable*}
%\end{document}

\subsection{HST follow-up with STIS} \label{sec:stis}

We obtained STIS spatially-resolved spectroscopy in the G750L mode covering a wavelength range from 5240-10270 \AA\ (dot-dash line in Figure \ref{fig:discovery}; Cycle 29, PID 16794). The $52\arcsec \times 0\farcs2$ slit was oriented at a position angle of $177.286^{\circ}$ to capture both components in a single observation. 
The STIS CCD has a plate scale of 0.05078”/pixel such that the two components are separated by $\sim 5$ pixels. 
The standard STIS reduction pipeline was used to remove detector signatures and defringe the spectra using the STIStools \texttt{defringe.defringe} command to remove the fringing pattern.

We use the \texttt{x1d} routine to extract each spectrum, adjusting the parameters to minimize overlap between the two. 
We constrain the search region for finding a peak in the extraction profile by setting \texttt{MAXSRCH} to 1.5 for each source and \texttt{A2CENTER} to 506 and 511 pixels, respectively. 
We set the extraction box size to 3 pixels and use a 10 pixel offset from the peak, which is far from both source profiles, for the background subtraction region. 
Two distinct spectra were extracted at 511.383 pixels and 506.324 pixels, respectively. 

\begin{figure}[ht!]
\centering
\includegraphics[width=\linewidth]{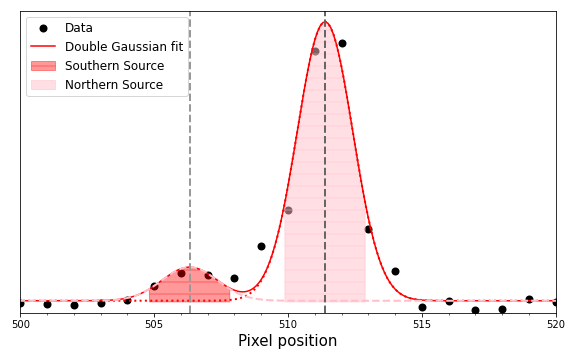}
\caption{Spatial profile of the source spectrum collapsed along the x-direction. Black points are the summed counts at each spatial pixel position, and the red line is a double Gaussian fit to the data. The shaded areas represent our 3\arcsec\ extraction regions, chosen to minimize blending.
Two distinct peaks are shown with the southern spectrum overlapping the northern spectrum by $\sim 0.6$\%. 
\label{fig:profile}}
\end{figure}

To evaluate the the impact of blending on our spectral extraction, we sum the cosmic-ray cleaned, normalized, and defringed science spectrum along the wavelength axis and plot the spatial profile of the two spectra in Figure \ref{fig:profile}. 
We fit two Gaussian distributions to this summed profile, keeping the width of the Gaussians tied to each other, and fixing the centers to the positions found by \texttt{x1d}. 
The extracted region for each spectrum is shown in light and dark shaded pink. 
Using the best-fit $\sigma$ of 1.04 pixels, we calculate that the southern spectrum overlaps the northern spectrum by $\sim 0.6$\%, ensuring that the individual spectra are not contaminated by blending.
We also determine that the 3-pixel aperture loses 14.8\% of the total flux. We correct the fluxes of our spectra by this amount. 

\begin{figure*}[ht!]
\centering
\includegraphics[width=\linewidth]{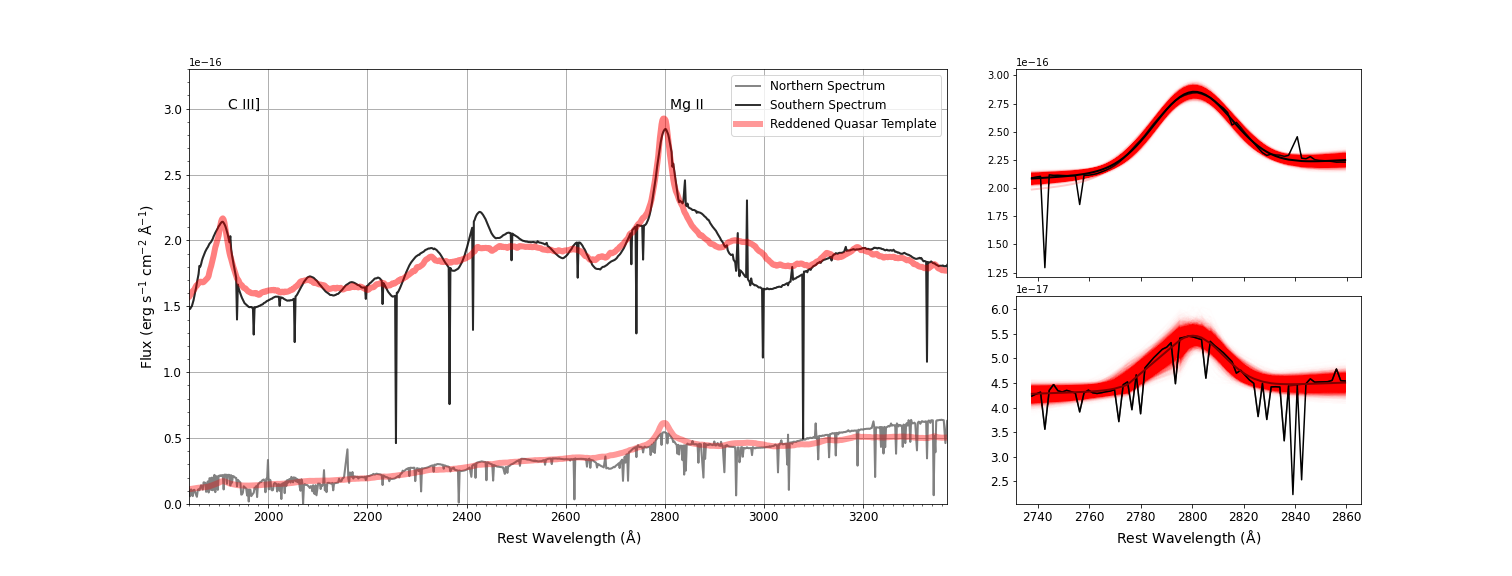}
\caption{{\em Left  -- } Individual spectra of the two QSO components plotted at rest wavelengths. The pink curves represent the best-fit reddened QSO template. The northern source (gray line) is reddened by $E(B-V) = 0.432$ while the southern source (black line) is reddened by $E(B-V) = 0.184$. \ion{Mg}{2} and \ion{C}{3} lines are labeled.
{\em Right -- } Gaussian fits to the \ion{Mg}{2} emission line in the southern (top) and northern (bottom) spectra showing 10,000 iterations determined by perturbing the best-fit line using the error arrays. The range of fits reflects the uncertainty in the derived Gaussian parameters. 
\label{fig:stis}}
\end{figure*}

Figure \ref{fig:stis}, left, shows the resultant extracted spectra for both QSO components. The \ion{Mg}{2} line that is seen in the SDSS spectrum (Fig. \ref{fig:discovery}) is visible in both the southern and northern components at $2800$\AA. However, the \ion{C}{3}] line at $\sim 2000$\AA\ is only seen in the southern component, where the signal-to-noise ratio is sufficiently high. The two spectra have different continuum shapes and the redder color of the northern component seen in the WFC3/IR image is apparent in the spectrum as well. 

\subsection{VLBA Imaging} \label{sec:vlba}

W2M~J1220 is detected in the FIRST catalog \citep{Becker03} with a 20 cm integrated flux density of 2.33 mJy. The peak flux density is 1.50 mJy/beam with an rms of 0.146 mJy/beam. The source's deconvolved major and minor axes are 5\farcs57 and 2\farcs26, respectively, indicating that the image is slightly resolved\footnote{The FIRST survey has an angular resolution of 5\arcsec.}. 

Aiming to detect two distinct radio components at their optical positions, we obtained 257 minutes of on-source integration with the Very Long Baseline Array (VLBA) split into two equal-length dual polarization observations on 19 August 2021 and 03 December 2021 in the L-band (1.4 GHz or 20 cm). We used J1218+1105 as a phase calibrator, which we measure to have an integrated flux density of 0.177 Jy in L-band located only 0.58 deg away from our target. 

\begin{figure}[ht!]
\centering
\includegraphics[width=\linewidth]{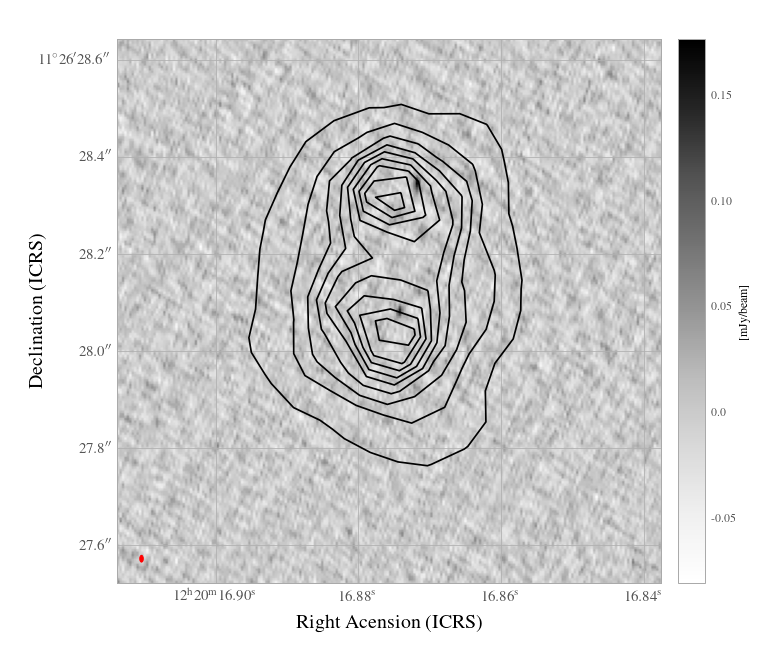}
\caption{Calibrated and cleaned image of VLBA L-band observations of W2M~J1220 produced with CASA. Overplotted contours indicate the flux from the WFC3/IR F160W image at 5$\sigma$ levels. Two point sources are detected at the HST position with a separation of 0\farcs26 and a position angle of 172.205$^{\circ}$. Beam size of 0\farcs005 is shown in red in the bottom left. 
\label{fig:vlba}}
\end{figure}

The observations were flagged, calibrated, cleaned, and imaged with the Common Astronomy Software Applications \citep[CASA;][]{CASA} package Version 6.5, following the approach described in VLBA Science Memo \#38 \citep{Linford22}. Fort Davis (FD) was selected as the reference antenna. To ensure that the amplitude scaling accounted for the wide bandpass, the task \texttt{ACCOR} was run twice – first for the initial calibrations and again after the bandpass correction. A phase-referenced (Stokes I) image of the target was produced by applying the \texttt{TCLEAN} task with natural weighting. The final calibrated image spans 320 x 0.001" along each axis with an rms noise level of 0.017 mJy and is shown in Figure \ref{fig:vlba}. 

The calibrated VLBA image shows two distinct point sources oriented at a position angle of 172.205$^{\circ}$ with a separation of 0\farcs26 (2.2 kpc). We performed 2D Gaussian fitting on each source with CASAviewer and found that the northern source has an integrated flux density of 0.502 $\pm$ 0.066 mJy and a peak flux density of 0.165 $\pm$ 0.016 mJy/beam. The deconvolved major and minor axis are 0\farcs0234 and 0\farcs0087. The southern source has an integrated flux density of 0.330 $\pm$ 0.044 mJy, a peak flux density of 0.146 $\pm$ 0.014 mJy/beam, and deconvolved major and minor axes of 0\farcs0166 and 0\farcs0072. 
In both sources, the major axis is slightly larger than the CLEAN beam which has FWHM of 0\farcs01 along both axes.

We overlay in contours the HST WFC3/IR F160W fluxes. 
The two VLBA point sources and their position angles are consistent with the HST position, though outside the 0\farcs03 Gaia-based astrometric errors, confirming their associations as the two components of W2M~J1220.

\section{Results} \label{sec:qso}

With our near-infrared imaging, spatially-resolved optical spectroscopy, and 20 cm radio imaging in hand, we are able to analyze the properties of the individual QSO components. We noticed that the SDSS-assigned redshift of $z=1.871$ was based on the \ion{C}{4} line, which did not align with the \ion{Mg}{2} line in the STIS spectra. Since \ion{C}{4} is known to be blushifted relative to QSOs' systemic redshifts \citep{Richards11}, we update the source redshift to $z=1.889$ based on the \ion{Mg}{2} line center derived from these new observations. 
We fit a reddened QSO template to each spectrum and find that the southern source has $E(B-V)=0.179\pm0.001$ while the northern source has $E(B-V) = 0.458\pm0.009$. These fits are shown with a pink line in Figure \ref{fig:stis}.  We correct the observed F160W photometry, corresponding to rest-fame 5320\AA, by these extinction values and, applying a bolometric correction of 9.2 \citep{Richards06}, compute $L_{\rm bol,south}=3.06\times10^{44}$ erg s$^{-1}$ and $L_{\rm bol,north}=4.84\times10^{44}$ erg s$^{-1}$.  This means that the northern source, which appears fainter, is intrinsically more luminous after correcting for its substantially higher amount of extinction.  We note that the intrinsically more luminous component coincides with the brighter radio source. 

We fit a Gaussian profile to the \ion{Mg}{2} line in both spectra to measure the $v_{\rm FWHM}$ values as $2830\pm650$ km s$^{-1}$ and $3920\pm260$ km s$^{-1}$ in the northern and southern sources, respectively. The errors are computed by perturbing the best fit model using the spectrum's error array and re-fitting 10,000 times. We compute the standard deviation of the Gaussian $\sigma$ parameter found in each fit iteration, shown in the right panels Figure \ref{fig:stis}. 

For the two QSOs, we apply the single-epoch virial black hole mass estimator ($M_{\rm BH}$) following the formalism of \citet{Shen12},
\begin{equation}
\log \bigg(\frac{M_{\rm BH,vir}}{M_\odot} \bigg) = a + b \log \bigg(\frac{L_{3000}}{10^{44} \rm erg/s} \bigg) + c \log \bigg(\frac{v_{\rm FWHM}}{\rm km/s} \bigg),
\label{eqn:mbh}
\end{equation}
adopting the values $a=0.740$, $b=0.620$, $c=2.00$ for single-epoch measurements of FWHM$_{\rm MgII}$ and $L_{3000}$, based on the calibration of \citet{Shen11}. For this calculation, we estimate $L_{3000}$ two different ways. We measure it directly from the STIS spectra by applying an aperture correction to the observed flux, and de-reddening each spectrum. We then apply an artificial 30 \AA-wide box-car filter centered on 3000 \AA\ to measure the source flux, from which we determine luminosity. The second method starts with the F160W source magnitudes (Table \ref{tab:phot}), which are far less sensitive to uncertainties in $E(B-V)$. We de-redden these magnitudes and use spectrophotometry to scale a QSO composite template to match the F160W flux. From the scaled template, we measure the 3000 \AA\ flux using the box-car filter as in the previous method. This results in a range of black hole masses, listed in Table \ref{tab:prop}. The BH masses differ for each method by $\lesssim 0.5$ dex but are extremely similar ($\lesssim 0.2$ dex) between the two components. These masses are at the high end of the range accessible to LISA.

Combining $M_{BH}$ and $L_{\rm bol}$ allows us to estimate the Eddington ratio. We find that the more obscured northern source has $L/L_{\rm Edd} \simeq 0.1 - 0.3$, while the less obscured, southern source has $L/L_{\rm Edd} \simeq 0.04 - 0.1$. This is consistent with findings that red QSOs have higher accretion rates than their unobscured counterparts \citep{Urrutia12,Kim15}. 

From the definition, $R \equiv f(1.4GHz) / f(B)$, we calculated the radio loudness parameter of the two sources. The optical flux is determined by the method described above, where the QSO template, scaled to the de-reddened F160W flux, is passed through a Johnson $B$ filter curve. 
The radio flux is not K-corrected, given that the radio spectral index for each source is not known.
We find that both sources have nearly identical $R$ values, at the boundary of the radio-quiet regime\footnote{Objects with $R>$ 2 are categorized as "radio-loud", while objects are generally considered "radio-quiet" when $R <$ 0.5. Radio-intermediate sources are those that fit neither category 0.5 $<R<$ 2 \citep{Stocke92}.}, with $R_{north}\approx 0.46$ and $R_{south}\approx$ 0.48.
Table \ref{tab:prop} lists all the derived properties for the two QSOs in this dual system.

%% LaTeX deluxetable generator for the AASTeX package.
%% Written by Greg Schwarz (5/1/2001).

%% Table generated: Thu Jul 28 16:19:28 2022

%% Remove the two lines and the last line if you want
%% want to incorporate this table into another LaTex document.
%\documentclass{aastex}
%\begin{document}

%% The values (usually only l,r and c) in the last part of
%% \begin{deluxetable}{} command tell LaTeX how many columns
%% there are and how to align them.
\begin{deluxetable*}{cccccccccc}

%% Rotate to a landscape orientation
%\rotate

%% Over-ride the default font size
%% Use Default (12pt)

%% Use \tablewidth{?pt} to over-ride the default table width.
%% If you are unhappy with the default look at the end of the
%% *.log file to see what the default was set at before adjusting
%% this value.
%% This is the title of the table.
\tablecaption{Individual QSO characteristics} 

%% This command over-rides LaTeX's natural table count
%% and replaces it with this number.  LaTeX will increment 
%% all other tables after this table based on this number
%\tablenum{2}

%% The \tablehead gives provides the column headers.  It
%% is currently set up so that the column labels are on the
%% top line and the units surrounded by ()s are in the 
%% bottom line.  You may add more header information by writing
%% another line between these lines. For each column that requries
%% extra information be sure to include a \colhead{text} command
%% and remember to end any extra lines with \\ and include the 
%% correct number of &s.
\tablehead{\colhead{Source} & \colhead{R.A.\tablenotemark{a}} & \colhead{Decl.\tablenotemark{a} } & \colhead{$E(B-V)$} & \colhead{$v_{\rm FWHM}$} & \colhead{$\log{M_{\rm BH}}$\tablenotemark{b}} & \colhead{$\log{L_{\rm bol}}$} & \colhead{$L/L_{\rm Edd}$} & \colhead{$S_{\rm pk,20cm}$} & \colhead{$R$} \\ 
\colhead{} & \colhead{(J2000)} & \colhead{(J2000)} & \colhead{(mag)} & \colhead{(km s$^{-1}$)} & \colhead{($M_\odot$)} & \colhead{(erg s$^{-1}$)} & \colhead{} & \colhead{(mJy)} & \colhead{} } 

%% All data must appear between the \startdata and \enddata commands
\startdata
North & 12:20:16.87176 & +11:26:28.344 & $0.458\pm0.009$ & $3140\pm800$ & $(7.2 - 7.7)\pm 0.2$    & 44.76 & $0.08 - 0.22$ & $0.502\pm0.066$ & $0.460$ \\
South & 12:20:16.87420 & +11:26:28.082 & $0.179\pm0.001$ & $3800\pm230$ & $(7.40 - 7.70)\pm 0.04$ & 44.46 & $0.05 - 0.1$ & $0.330\pm0.044$ & $0.478$ \\
\enddata

%% Include any \tablenotetext{key}{text}, \tablerefs{ref list},
\tablenotetext{a}{The celestial coordinate positions are from the VLBA data and have the uncertainties $\delta_{\rm R.A.} = 0.00003$s and $\delta_{\rm Decl.} = 0.001$\arcsec.} 
\tablenotetext{b}{The range of $\log{M_{\rm BH}}$ is based on the two methods for estimating $L_{3000}$ while the quoted uncertainties are based on the propagation of the $v_{\rm FWHM}$ uncertainties and are therefore lower limits.} 
%% between the \enddata and \end{deluxetable} commands

%% No \tablecomments indicated

%% No \tablerefs indicated
\label{tab:prop}

\end{deluxetable*}
%\end{document}

\section{Discussion} \label{sec:floats}

The confirmation of two distinct QSO spectra, at the same redshift, separated by 0\farcs26, and coincident with two compact radio sources provides strong evidence that W2M~J1220 is a dual QSO. 
Most of the known and confirmed dual AGNs are at low redshifts \citep[$z < 0.7$; e.g.][]{Koss12,Comerford12,MullerSanchez15,Fu15,Rubinur19}, which is not yet probing the epoch of peak QSO activity in the universe \cite[$z\simeq2$;][]{Madau14} when merger rates were significantly higher \citep{Conselice03,RodriguezGomez15}. 
Therefore, the identification of a dual QSO system at this epoch is noteworthy, especially given that red QSOs are predominantly found in merging hosts.

W2M~J1220 is comparable to LBQS~0103$-$2753 \citep{Shields12}, which is a confirmed dual QSO, separated by 0\farcs3, at $z=0.858$. LBQS 0103$-$2753 was identified in HST imaging and verified with a STIS spectrum, similar to W2M~J1220. Deep HST imaging of LBQS 0103$-$2753 reveals tidal features and morphological evidence of a recent merger. The WFC3/IR imaging for W2M~J1220 is not deep enough to show these features. The two-component spectra of LBQS 0103$-$2753 are quite distinct, with one of the components showing BAL features indicative of outflows. There is also a velocity offset of $\sim 1500$ km s$^{-1}$ between the two components. Although the black hole masses of LBQS 0103$-$2753 are $\sim 1-1.5$ orders of magnitude higher than W2M~J1220, they are similar to each other (both have $M_{\rm BH} \sim 10^{8.5 - 9}~M_\odot$).

In the cosmic noon era ($z\sim 2-3$), \citet{Shen21} report two dual QSO candidates using the novel technique of `varstrometry' which identifies sources with high astrometric variability in Gaia suggestive of two distinct, closely-spaced sources with randomly varying fluxes \citep{Hwang20}. One source, J0841+4825, is at $z=2.95$ and is separated by 0\farcs46, though its ground-based spatially-resolved spectroscopy shows highly similar spectra which could be explained by gravitational lensing. Both components of J0749+2255, at $z=2.17$ and also separated by 0\farcs46, are detected by VLBA observations at 15 GHz, has a spatially resolved STIS spectrum, and HST imaging showing merger signatures in the host, putting this system on solid footing for a dual QSO \citep{Chen23}.
There are 45 additional varstrometry-selected dual QSO candidates extending out to $z\simeq3$ awaiting confirmation \citep{Chen22}.

The gravitational lens PSJ1721+8842, initially thought to be a quadruple lens at $z=2.37$, was analyzed by \citet{Mangat21} to re-interpret the system as two QSOs that are lensed to form four point source images based on HST optical and IR observations as well as VLA observations. 

\citet{Yue21} report a candidate QSO pair at $z=5.66$ separated by 1\farcs24, or 7.3 kpc. Spatially resolved spectroscopy reveal two spectra with similar line characteristics but different reddenings, as in W2M~J1220. 

\subsection{Lensing considerations}

Some of the properties between the two components of W2M~J1220, such as the derived black hole masses, radio loudness parameters, as well as near-identical emission line centers and profile shapes, may be explained by gravitational lensing. Here we consider that possibility. 

\begin{figure}
\includegraphics[width=\linewidth]{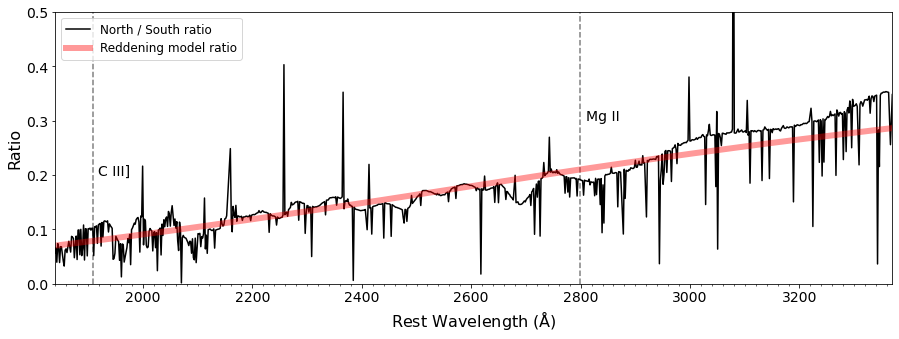}
\caption{Flux ratio of the northern to southern spectra showing the disappearance of the \ion{C}{4} and \ion{Mg}{2} emission lines (vertical dashed lines), motivating an exploration of gravitational lensing as the cause for the pair of QSO images. The pink line represents a ratio of the best-fit reddening curves shown in Figure \ref{fig:stis}.}
\label{fig:ratio}
\end{figure}

Figure \ref{fig:ratio} shows the ratio of the northern to southern spectrum, with the ratio of the best-fit reddened templates (pink curves in Figure \ref{fig:stis}) over-plotted. While the \ion{C}{4} and \ion{Mg}{2} emission lines, marked by vertical dashed lines, disappear, the shape of the ratio is broadly consistent with the difference in reddening.
The undulating features that deviate from a QSO template seen in both the ratio spectrum and in the southern component are not consistent with known QSO spectral features, such as iron emission in the ultraviolet \citep[UV;][]{Vestergaard01}.

With a separation of 0\farcs26, corresponding to an Einstein radius of 0\farcs13, W2M~J1220 would be the most closely-separated lensed QSO known\footnote{J0439+1634 has a separation of 0\farcs2 \citep{Fan19}. This source is of the rare class of “naked” cusp lenses involving three images, which is ruled out for our system by the VLBA image. B0218+357, with a separation of 0\farcs34 is currently the next most closely-separated lens \citep{Patnaik93}.} Such small-separation lenses \citep[on the order of $\sim 100$ milli-arcseconds or, so-called, milli-lenses;][]{Frey10,Spingola19,Casadio21} are rare and have been difficult to find, but probe supermassive ($10^6 - 10^9$) compact objects as putative lenses. Recent systematic searches using VLBA data have resulted in few viable candidates. This is because a lens with an unusually high surface density is needed to yield such small separations. 

We explore the range of possible lens masses that could result in a separation of 0\farcs26 as a function of redshift to determine the plausibility of lensing in W2M1220.
We employ the relation among distances, lens mass, and source separation, and a point mass lens, 
\begin{equation}
    M(\theta_E) = \left( \frac{c^2}{4G} \right) \theta_E^2 \left( \frac{D_d  D_s}{D_{ds}}\right),
\end{equation}
and for a singular isothermal sphere (SIS) mass model,
\begin{equation}
    M(\theta_E) = \frac{\sigma_v^2 R}{G} ~ \rm{ ~with ~} ~ \sigma_v = c \sqrt{\frac{\theta_E}{4\pi} \frac{D_s}{D_{ds}} }.
\end{equation}
Here, $M(\theta_E)$ is the mass enclosed within the angular radius $\theta_E$, $D_d$ and $D_s$ are the angular diameter distance to the lens, source, respectively, and $D_{ds}$ is the angular diameter distance between the lens and source. Given the redshift of W2M~J1220, $D_s$ is known we can compute $M(\theta_E)$ as a function of lens redshift. We find that even the smallest lens masses, residing at $z \sim 0.4 - 0.6$, would require $>10^9~M_\odot$ to be confined to the innermost kpc$^2$. 

Assuming lensing as a possibility, we model the system with GRAVLENS \citep{Keeton01} and determine the lens position that would yield the two VLBA positions finding that a lens would need to be situated 0\farcs021695 to the west and 0\farcs159543 to the south of the northern source. An attempt to fix a S\'{e}rsic component at this position relative to fixed PSF components in our morphological fitting (\S \ref{sec:discovery}) resulted in failure of Galfit to converge. In fact, the central S\'{e}rsic component found by Galfit, while possessing unphysical properties, is situated to the {\em east} of both PSF components.

Identifying intrinsic differences, such as in the spectral slope, between the two QSOs would rule out lensing.
The $E(B-V)$ values that we find are determined with respect to a composite QSO spectrum that has a spectral slope of $\alpha_\nu = -0.47$ at $\lambda < 5000$. 
We vary the slope of the QSO template \citep[following Section 5.3 of][]{Glikman07} and recomputed $E(B-V)$ for a template with $\alpha_\nu = -0.25$ (a bluer slope) and $\alpha_\nu = -0.76$ (a redder slope), which represent the range intrinsic to unreddened QSOs \citep{Richards03}. 
While the bluer template does yield slightly higher values by $\Delta E(B-V)\sim0.02$ (and vice-versa for the redder template, yielding lower values by $\Delta E(B-V)\sim0.03$), the differences are not sufficient to account for the difference in $E(B-V)$ between the two sources as seen in the continuum fits (Figure \ref{fig:stis}), the flux ratios (Figure \ref{fig:ratio}) and the $F105W - F160W$ colors.
We therefore cannot attribute the different reddenings to intrinsic differences in spectral slopes between the two spectra and the best explanation for the $E(B-V)$ values is different amounts of dust reddening along the two lines of sight.

To achieve the observed amount of reddening would require a lensing galaxy with a significant amount of gas and dust.
While we cannot rule out lensing with this spectral slope investigation, it does rule out the exotic possibility that the lens might be a freely floating SMBH.

Finally, the similarities between the black hole properties in W2M~J1220 are seen in previously confirmed dual QSOs, such as LBQS 0103$-$2753 \citep{Shields12}. Simulations of SMBH binaries predict major mergers more often produce dual AGNs with similar BH masses \citep{Blecha13,Steinborn16}.
And, since red QSOs are known to be associated with major mergers, this discovery may reflect a selection effect towards similar mass BHs.

The similarity in radio loudness may reflect the enhanced low-level radio emission seen in W2M red QSOs, which has been interpreted as coming from either a dusty wind or nascent jets \citep{Glikman22}. 
Likewise, the differences between the two QSOs, such as the higher accretion rate seen in the redder source, is consistent with what has been seen in red QSOs elsewhere \citep{Urrutia12,Kim15}.

Finally, as some gravitational lenses show intervening absorption features in the individual component spectra \citep[e.g.,][]{Rubin18}, which may reveal the putative lens redshift, we explored the absorption features in the STIS spectra and could not identify any evidence for such coherent features. This further weakens the possibility of a gas rich galaxy as a gravitational lens.
Therefore, although we cannot definitively rule it out, we consider lensing to be the less likely explanation for this system. 

One way to rule out lensing would be to obtain deeper imaging with HST that may reveal evidence of a merging system, as is seen in the dual QSO found by \citet{Chen23}. Another way would be to measure the flux densities at other radio frequencies and compare their radio spectral indices which, if different for each source, would also rule out lensing.  
And, if a lens is responsible, its nature as highly compact and with extreme dust gradients, is worthy of its own study.

\subsection{dual statistics for red QSOs}\label{sec:stat}

The serendipitous discovery of a QSO pair in HST imaging of a red QSO raises the question of their frequency compared to unreddened QSOs. 
\citet{Shen23} investigate statistically the incidence of QSO pairs using Gaia detections of known SDSS QSOs with $L_{\rm bol} > 10^{45.8}$ erg s$^{-1}$ at $1.5<z<3.5$ with separations of $0\farcs4 - 3\arcsec$ and find an integrated pair fraction of $\sim 6\times 10^{-4}$. 
Assuming this fraction is constant at the 0\farcs26 separation of W2M~1220 and is a factor of $\sim 10$ higher given its lower luminosity ($L_{\rm bol} \sim 10^{44.8}$ erg s$^{-1}$), the pair fraction is estimated to be $\sim 10^{-3}$ \citep{Shen23}. However, only 17 red QSOs have been imaged with HST at $z\sim2$ \citep[11 F2M quasars reported in][and 6 W2M QSOs which include W2M~1220]{Glikman15}. This fraction of 0.06 (1/17) would be an order of magnitude higher than that found for the luminous, unobscured QSOs investigated in \citet{Shen23}. 
Given that red QSOs are known to be hosted by major mergers, this population may be the most likely for finding dual QSOs although with only a single system, we cannot draw broad conclusions.

\section{conclusions} \label{sec:conclusions}

We report the discovery of a dual QSO candidate, separated by 0\farcs26 corresponding to 2.2 kpc at $z=1.889$. The sources are confirmed as QSOs with a spatially-resolved STIS spectrum and high-resolution VLBA imaging at 1.4 GHz which reveal two point sources consistent with the positions in the HST images. The two components are reddened by different amounts of dust-extinction. When corrected for this extinction, the properties of the QSOs are similar, including black hole masses $\sim 10^{7.5} M_\odot$ and radio loudness of $\sim0.5$ (though their Eddington ratios differ). These similarities mean we cannot rule out gravitational lensing, though the lens is not detected in the imaging and extended features seen in the HST imaging may indicate merging hosts. 
The features of these two QSOs are consistent with previous findings in dual AGNs.

A dual QSO discovered at cosmic noon in a survey for red QSOs, which is a population known to be hosted by major mergers, can provide a unique population in which to search for such systems where both black holes are active at the same time. 
Given that only $\sim 30$ red quasars have been observed with HST, finding a candidate dual QSO in such a small sample suggests an elevated incidence of dual activity in red QSOs. 
Because W2M~J1220 was found serendipitously, a targeted high resolution imaging effort of red QSOs at $z=2-3$ may be the most fruitful place to find dual quasars during a crucial phase of SMBH/Galaxy co-evolution.

\begin{acknowledgments}
We thank Marianne Vestergaard for sharing the Fe UV template which we used to look for features in our spectra. 
E.G. acknowledges the generous support of the Cottrell Scholar Award through the Research Corporation for Science Advancement. 
E.G. is grateful to the Mittelman Family Foundation for their generous support. 
We gratefully acknowledge the National Science Foundation's support of the Keck Northeast Astronomy Consortium's REU program through grant AST-1950797.   
BDS acknowledges support through a UK Research and Innovation Future Leaders Fellowship [grant number MR/T044136/1].

Some/all of the data presented in this paper were obtained from the Mikulski Archive for Space Telescopes (MAST) at the Space Telescope Science Institute. The specific observations analyzed can be accessed via \dataset[10.17909/5ydb-ex84]{https://doi.org/10.17909/5ydb-ex84} and \dataset[10.17909/s2sz-t252]{https://doi.org/10.17909/s2sz-t252}. 

This research is based on observations made with the NASA/ESA {\em Hubble} Space Telescope obtained from the Space Telescope Science Institute, which is operated by the Association of Universities for Research in Astronomy, Inc., under NASA contract NAS 5–26555. These observations are associated with program(s) PID 16794.

The National Radio Astronomy Observatory is a facility of the National Science Foundation operated under cooperative agreement by Associated Universities, Inc.  
This work made use of the Swinburne University of Technology software correlator \citep{Deller11}, developed as part of the Australian Major National Research Facilities Programme and operated under licence.

\end{acknowledgments}

\vspace{5mm}
\facilities{HST(STIS), VLBA, SDSS, Palomar(TripleSpec)}

\software{astropy \citep{2013A&A...558A..33A,2018AJ....156..123A},  
          astroconda,
          CASA
          }
          
\dataset[10.17909/5ydb-ex84]{http://dx.doi.org/10.17909/5ydb-ex84}
\dataset[10.17909/s2sz-t252]{http://dx.doi.org/10.17909/s2sz-t252}

\bibliography{dualQSO.bbl}

\begin{thebibliography}{}
\expandafter\ifx\csname natexlab\endcsname\relax\def\natexlab#1{#1}\fi
\providecommand{\url}[1]{\href{#1}{#1}}
\providecommand{\dodoi}[1]{doi:~\href{http://doi.org/#1}{\nolinkurl{#1}}}
\providecommand{\doeprint}[1]{\href{http://ascl.net/#1}{\nolinkurl{http://ascl.net/#1}}}
\providecommand{\doarXiv}[1]{\href{https://arxiv.org/abs/#1}{\nolinkurl{https://arxiv.org/abs/#1}}}

\bibitem[{{Astropy Collaboration} {et~al.}(2013){Astropy Collaboration},
  {Robitaille}, {Tollerud}, {Greenfield}, {Droettboom}, {Bray}, {Aldcroft},
  {Davis}, {Ginsburg}, {Price-Whelan}, {Kerzendorf}, {Conley}, {Crighton},
  {Barbary}, {Muna}, {Ferguson}, {Grollier}, {Parikh}, {Nair}, {Unther},
  {Deil}, {Woillez}, {Conseil}, {Kramer}, {Turner}, {Singer}, {Fox}, {Weaver},
  {Zabalza}, {Edwards}, {Azalee Bostroem}, {Burke}, {Casey}, {Crawford},
  {Dencheva}, {Ely}, {Jenness}, {Labrie}, {Lim}, {Pierfederici}, {Pontzen},
  {Ptak}, {Refsdal}, {Servillat}, \& {Streicher}}]{2013A&A...558A..33A}
{Astropy Collaboration}, {Robitaille}, T.~P., {Tollerud}, E.~J., {et~al.} 2013,
  \aap, 558, A33, \dodoi{10.1051/0004-6361/201322068}

\bibitem[{{Astropy Collaboration} {et~al.}(2018){Astropy Collaboration},
  {Price-Whelan}, {Sip{\H{o}}cz}, {G{\"u}nther}, {Lim}, {Crawford}, {Conseil},
  {Shupe}, {Craig}, {Dencheva}, {Ginsburg}, {VanderPlas}, {Bradley},
  {P{\'e}rez-Su{\'a}rez}, {de Val-Borro}, {Aldcroft}, {Cruz}, {Robitaille},
  {Tollerud}, {Ardelean}, {Babej}, {Bach}, {Bachetti}, {Bakanov}, {Bamford},
  {Barentsen}, {Barmby}, {Baumbach}, {Berry}, {Biscani}, {Boquien}, {Bostroem},
  {Bouma}, {Brammer}, {Bray}, {Breytenbach}, {Buddelmeijer}, {Burke},
  {Calderone}, {Cano Rodr{\'\i}guez}, {Cara}, {Cardoso}, {Cheedella}, {Copin},
  {Corrales}, {Crichton}, {D'Avella}, {Deil}, {Depagne}, {Dietrich}, {Donath},
  {Droettboom}, {Earl}, {Erben}, {Fabbro}, {Ferreira}, {Finethy}, {Fox},
  {Garrison}, {Gibbons}, {Goldstein}, {Gommers}, {Greco}, {Greenfield},
  {Groener}, {Grollier}, {Hagen}, {Hirst}, {Homeier}, {Horton}, {Hosseinzadeh},
  {Hu}, {Hunkeler}, {Ivezi{\'c}}, {Jain}, {Jenness}, {Kanarek}, {Kendrew},
  {Kern}, {Kerzendorf}, {Khvalko}, {King}, {Kirkby}, {Kulkarni}, {Kumar},
  {Lee}, {Lenz}, {Littlefair}, {Ma}, {Macleod}, {Mastropietro}, {McCully},
  {Montagnac}, {Morris}, {Mueller}, {Mumford}, {Muna}, {Murphy}, {Nelson},
  {Nguyen}, {Ninan}, {N{\"o}the}, {Ogaz}, {Oh}, {Parejko}, {Parley}, {Pascual},
  {Patil}, {Patil}, {Plunkett}, {Prochaska}, {Rastogi}, {Reddy Janga},
  {Sabater}, {Sakurikar}, {Seifert}, {Sherbert}, {Sherwood-Taylor}, {Shih},
  {Sick}, {Silbiger}, {Singanamalla}, {Singer}, {Sladen}, {Sooley},
  {Sornarajah}, {Streicher}, {Teuben}, {Thomas}, {Tremblay}, {Turner},
  {Terr{\'o}n}, {van Kerkwijk}, {de la Vega}, {Watkins}, {Weaver}, {Whitmore},
  {Woillez}, {Zabalza}, \& {Astropy Contributors}}]{2018AJ....156..123A}
{Astropy Collaboration}, {Price-Whelan}, A.~M., {Sip{\H{o}}cz}, B.~M., {et~al.}
  2018, \aj, 156, 123, \dodoi{10.3847/1538-3881/aabc4f}

\bibitem[{{Becker} {et~al.}(2003){Becker}, {Helfand}, {White}, {Gregg}, \&
  {Laurent-Muehleisen}}]{Becker03}
{Becker}, R.~H., {Helfand}, D.~J., {White}, R.~L., {Gregg}, M.~D., \&
  {Laurent-Muehleisen}, S.~A. 2003, VizieR Online Data Catalog, VIII/71

\bibitem[{{Blecha} {et~al.}(2013){Blecha}, {Loeb}, \& {Narayan}}]{Blecha13}
{Blecha}, L., {Loeb}, A., \& {Narayan}, R. 2013, \mnras, 429, 2594,
  \dodoi{10.1093/mnras/sts533}

\bibitem[{{CASA Team} {et~al.}(2022){CASA Team}, {Bean}, {Bhatnagar}, {Castro},
  {Donovan Meyer}, {Emonts}, {Garcia}, {Garwood}, {Golap}, {Gonzalez Villalba},
  {Harris}, {Hayashi}, {Hoskins}, {Hsieh}, {Jagannathan}, {Kawasaki},
  {Keimpema}, {Kettenis}, {Lopez}, {Marvil}, {Masters}, {McNichols},
  {Mehringer}, {Miel}, {Moellenbrock}, {Montesino}, {Nakazato}, {Ott}, {Petry},
  {Pokorny}, {Raba}, {Rau}, {Schiebel}, {Schweighart}, {Sekhar}, {Shimada},
  {Small}, {Steeb}, {Sugimoto}, {Suoranta}, {Tsutsumi}, {van Bemmel},
  {Verkouter}, {Wells}, {Xiong}, {Szomoru}, {Griffith}, {Glendenning}, \&
  {Kern}}]{CASA}
{CASA Team}, {Bean}, B., {Bhatnagar}, S., {et~al.} 2022, \pasp, 134, 114501,
  \dodoi{10.1088/1538-3873/ac9642}

\bibitem[{{Casadio} {et~al.}(2021){Casadio}, {Blinov}, {Readhead}, {Browne},
  {Wilkinson}, {Hovatta}, {Mandarakas}, {Pavlidou}, {Tassis}, {Vedantham},
  {Zensus}, {Diamantopoulos}, {Dolapsaki}, {Gkimisi}, {Kalaitzidakis},
  {Mastorakis}, {Nikolaou}, {Ntormousi}, {Pelgrims}, \& {Psarras}}]{Casadio21}
{Casadio}, C., {Blinov}, D., {Readhead}, A.~C.~S., {et~al.} 2021, \mnras, 507,
  L6, \dodoi{10.1093/mnrasl/slab082}

\bibitem[{{Chen} {et~al.}(2022){Chen}, {Hwang}, {Shen}, {Liu}, {Zakamska},
  {Yang}, \& {Li}}]{Chen22}
{Chen}, Y.-C., {Hwang}, H.-C., {Shen}, Y., {et~al.} 2022, \apj, 925, 162,
  \dodoi{10.3847/1538-4357/ac401b}

\bibitem[{{Chen} {et~al.}(2023){Chen}, {Liu}, {Foord}, {Shen}, {Oguri}, {Chen},
  {Di Matteo}, {Holgado}, {Hwang}, \& {Zakamska}}]{Chen23}
{Chen}, Y.-C., {Liu}, X., {Foord}, A., {et~al.} 2023, \nat, 616, 45,
  \dodoi{10.1038/s41586-023-05766-6}

\bibitem[{{Comerford} {et~al.}(2012){Comerford}, {Gerke}, {Stern}, {Cooper},
  {Weiner}, {Newman}, {Madsen}, \& {Barrows}}]{Comerford12}
{Comerford}, J.~M., {Gerke}, B.~F., {Stern}, D., {et~al.} 2012, \apj, 753, 42,
  \dodoi{10.1088/0004-637X/753/1/42}

\bibitem[{{Conselice} {et~al.}(2003){Conselice}, {Bershady}, {Dickinson}, \&
  {Papovich}}]{Conselice03}
{Conselice}, C.~J., {Bershady}, M.~A., {Dickinson}, M., \& {Papovich}, C. 2003,
  \aj, 126, 1183, \dodoi{10.1086/377318}

\bibitem[{{Deller} {et~al.}(2011){Deller}, {Brisken}, {Phillips}, {Morgan},
  {Alef}, {Cappallo}, {Middelberg}, {Romney}, {Rottmann}, {Tingay}, \&
  {Wayth}}]{Deller11}
{Deller}, A.~T., {Brisken}, W.~F., {Phillips}, C.~J., {et~al.} 2011, \pasp,
  123, 275, \dodoi{10.1086/658907}

\bibitem[{{Fan} {et~al.}(2019){Fan}, {Wang}, {Yang}, {Keeton}, {Yue},
  {Zabludoff}, {Bian}, {Bonaglia}, {Georgiev}, {Hennawi}, {Li}, {McGreer},
  {Naidu}, {Pacucci}, {Rabien}, {Thompson}, {Venemans}, {Walter}, {Wang}, \&
  {Wu}}]{Fan19}
{Fan}, X., {Wang}, F., {Yang}, J., {et~al.} 2019, \apjl, 870, L11,
  \dodoi{10.3847/2041-8213/aaeffe}

\bibitem[{{Fitzpatrick}(1999)}]{F99}
{Fitzpatrick}, E.~L. 1999, \pasp, 111, 63, \dodoi{10.1086/316293}

\bibitem[{{Frey} {et~al.}(2010){Frey}, {Paragi}, {Campbell}, \&
  {Mo{\'o}r}}]{Frey10}
{Frey}, S., {Paragi}, Z., {Campbell}, R.~M., \& {Mo{\'o}r}, A. 2010, \aap, 513,
  A18, \dodoi{10.1051/0004-6361/200913864}

\bibitem[{{Fu} {et~al.}(2015){Fu}, {Wrobel}, {Myers}, {Djorgovski}, \&
  {Yan}}]{Fu15}
{Fu}, H., {Wrobel}, J.~M., {Myers}, A.~D., {Djorgovski}, S.~G., \& {Yan}, L.
  2015, \apjl, 815, L6, \dodoi{10.1088/2041-8205/815/1/L6}

\bibitem[{{Gebhardt} {et~al.}(2000){Gebhardt}, {Bender}, {Bower}, {Dressler},
  {Faber}, {Filippenko}, {Green}, {Grillmair}, {Ho}, {Kormendy}, {Lauer},
  {Magorrian}, {Pinkney}, {Richstone}, \& {Tremaine}}]{Gebhardt00}
{Gebhardt}, K., {Bender}, R., {Bower}, G., {et~al.} 2000, \apjl, 539, L13,
  \dodoi{10.1086/312840}

\bibitem[{{Glikman}(2017)}]{Glikman17}
{Glikman}, E. 2017, Research Notes of the American Astronomical Society, 1, 48,
  \dodoi{10.3847/2515-5172/aaa0c0}

\bibitem[{{Glikman} {et~al.}(2004){Glikman}, {Gregg}, {Lacy}, {Helfand},
  {Becker}, \& {White}}]{Glikman04}
{Glikman}, E., {Gregg}, M.~D., {Lacy}, M., {et~al.} 2004, \apj, 607, 60

\bibitem[{{Glikman} {et~al.}(2006){Glikman}, {Helfand}, \& {White}}]{Glikman06}
{Glikman}, E., {Helfand}, D.~J., \& {White}, R.~L. 2006, \apj, 640, 579,
  \dodoi{10.1086/500098}

\bibitem[{{Glikman} {et~al.}(2007){Glikman}, {Helfand}, {White}, {Becker},
  {Gregg}, \& {Lacy}}]{Glikman07}
{Glikman}, E., {Helfand}, D.~J., {White}, R.~L., {et~al.} 2007, \apj, 667, 673,
  \dodoi{10.1086/521073}

\bibitem[{{Glikman} {et~al.}(2015){Glikman}, {Simmons}, {Mailly}, {Schawinski},
  {Urry}, \& {Lacy}}]{Glikman15}
{Glikman}, E., {Simmons}, B., {Mailly}, M., {et~al.} 2015, \apj, 806, 218,
  \dodoi{10.1088/0004-637X/806/2/218}

\bibitem[{{Glikman} {et~al.}(2012){Glikman}, {Urrutia}, {Lacy}, {Djorgovski},
  {Mahabal}, {Myers}, {Ross}, {Petitjean}, {Ge}, {Schneider}, \&
  {York}}]{Glikman12}
{Glikman}, E., {Urrutia}, T., {Lacy}, M., {et~al.} 2012, \apj, 757, 51,
  \dodoi{10.1088/0004-637X/757/1/51}

\bibitem[{{Glikman} {et~al.}(2018){Glikman}, {Lacy}, {LaMassa}, {Stern},
  {Djorgovski}, {Graham}, {Urrutia}, {Lovdal}, {Crnogorcevic}, {Daniels-Koch},
  {Hundal}, {Urry}, {Gates}, \& {Murray}}]{Glikman18}
{Glikman}, E., {Lacy}, M., {LaMassa}, S., {et~al.} 2018, \apj, 861, 37,
  \dodoi{10.3847/1538-4357/aac5d8}

\bibitem[{{Glikman} {et~al.}(2022){Glikman}, {Lacy}, {LaMassa}, {Bradley},
  {Djorgovski}, {Urrutia}, {Gates}, {Graham}, {Urry}, \& {Yoon}}]{Glikman22}
---. 2022, \apj, 934, 119, \dodoi{10.3847/1538-4357/ac6bee}

\bibitem[{{Gordon} \& {Clayton}(1998)}]{Gordon98}
{Gordon}, K.~D., \& {Clayton}, G.~C. 1998, \apj, 500, 816,
  \dodoi{10.1086/305774}

\bibitem[{{Hopkins} {et~al.}(2005){Hopkins}, {Hernquist}, {Cox}, {Di Matteo},
  {Martini}, {Robertson}, \& {Springel}}]{Hopkins05}
{Hopkins}, P.~F., {Hernquist}, L., {Cox}, T.~J., {et~al.} 2005, \apj, 630, 705,
  \dodoi{10.1086/432438}

\bibitem[{{Hopkins} {et~al.}(2008){Hopkins}, {Hernquist}, {Cox}, \&
  {Kere{\v{s}}}}]{Hopkins08}
{Hopkins}, P.~F., {Hernquist}, L., {Cox}, T.~J., \& {Kere{\v{s}}}, D. 2008,
  \apjs, 175, 356, \dodoi{10.1086/524362}

\bibitem[{{Hwang} {et~al.}(2020){Hwang}, {Shen}, {Zakamska}, \&
  {Liu}}]{Hwang20}
{Hwang}, H.-C., {Shen}, Y., {Zakamska}, N., \& {Liu}, X. 2020, \apj, 888, 73,
  \dodoi{10.3847/1538-4357/ab5c1a}

\bibitem[{{Keeton}(2001)}]{Keeton01}
{Keeton}, C.~R. 2001, arXiv e-prints, astro,
  \dodoi{10.48550/arXiv.astro-ph/0102340}

\bibitem[{{Kim} {et~al.}(2015){Kim}, {Im}, {Glikman}, {Woo}, \&
  {Urrutia}}]{Kim15}
{Kim}, D., {Im}, M., {Glikman}, E., {Woo}, J.-H., \& {Urrutia}, T. 2015, \apj,
  812, 66, \dodoi{10.1088/0004-637X/812/1/66}

\bibitem[{{Koss} {et~al.}(2012){Koss}, {Mushotzky}, {Treister}, {Veilleux},
  {Vasudevan}, \& {Trippe}}]{Koss12}
{Koss}, M., {Mushotzky}, R., {Treister}, E., {et~al.} 2012, \apjl, 746, L22,
  \dodoi{10.1088/2041-8205/746/2/L22}

\bibitem[{{Linford}(2022)}]{Linford22}
{Linford}, J. 2022, VLBA Scientific Memos, 38

\bibitem[{{Madau} \& {Dickinson}(2014)}]{Madau14}
{Madau}, P., \& {Dickinson}, M. 2014, \araa, 52, 415,
  \dodoi{10.1146/annurev-astro-081811-125615}

\bibitem[{{Magorrian} {et~al.}(1998){Magorrian}, {Tremaine}, {Richstone},
  {Bender}, {Bower}, {Dressler}, {Faber}, {Gebhardt}, {Green}, {Grillmair},
  {Kormendy}, \& {Lauer}}]{Magorrian98}
{Magorrian}, J., {Tremaine}, S., {Richstone}, D., {et~al.} 1998, \aj, 115,
  2285, \dodoi{10.1086/300353}

\bibitem[{{Mangat} {et~al.}(2021){Mangat}, {McKean}, {Brilenkov}, {Hartley},
  {Stacey}, {Vegetti}, \& {Wen}}]{Mangat21}
{Mangat}, C.~S., {McKean}, J.~P., {Brilenkov}, R., {et~al.} 2021, \mnras, 508,
  L64, \dodoi{10.1093/mnrasl/slab106}

\bibitem[{{Marconi} \& {Hunt}(2003)}]{Marconi03}
{Marconi}, A., \& {Hunt}, L.~K. 2003, \apjl, 589, L21, \dodoi{10.1086/375804}

\bibitem[{{M{\"u}ller-S{\'a}nchez} {et~al.}(2015){M{\"u}ller-S{\'a}nchez},
  {Comerford}, {Nevin}, {Barrows}, {Cooper}, \& {Greene}}]{MullerSanchez15}
{M{\"u}ller-S{\'a}nchez}, F., {Comerford}, J.~M., {Nevin}, R., {et~al.} 2015,
  \apj, 813, 103, \dodoi{10.1088/0004-637X/813/2/103}

\bibitem[{{Patnaik} {et~al.}(1993){Patnaik}, {Browne}, {King}, {Muxlow},
  {Walsh}, \& {Wilkinson}}]{Patnaik93}
{Patnaik}, A.~R., {Browne}, I.~W.~A., {King}, L.~J., {et~al.} 1993, \mnras,
  261, 435, \dodoi{10.1093/mnras/261.2.435}

\bibitem[{{Peng} {et~al.}(2002){Peng}, {Ho}, {Impey}, \& {Rix}}]{Peng02}
{Peng}, C.~Y., {Ho}, L.~C., {Impey}, C.~D., \& {Rix}, H.-W. 2002, \aj, 124,
  266, \dodoi{10.1086/340952}

\bibitem[{{Richards} {et~al.}(2003){Richards}, {Hall}, {Vanden Berk},
  {Strauss}, {Schneider}, {Weinstein}, {Reichard}, {York}, {Knapp}, {Fan},
  {Ivezi{\'c}}, {Brinkmann}, {Budav{\'a}ri}, {Csabai}, \&
  {Nichol}}]{Richards03}
{Richards}, G.~T., {Hall}, P.~B., {Vanden Berk}, D.~E., {et~al.} 2003, \aj,
  126, 1131, \dodoi{10.1086/377014}

\bibitem[{{Richards} {et~al.}(2006){Richards}, {Lacy}, {Storrie-Lombardi},
  {Hall}, {Gallagher}, {Hines}, {Fan}, {Papovich}, {Vanden Berk}, {Trammell},
  {Schneider}, {Vestergaard}, {York}, {Jester}, {Anderson}, {Budav{\'a}ri}, \&
  {Szalay}}]{Richards06}
{Richards}, G.~T., {Lacy}, M., {Storrie-Lombardi}, L.~J., {et~al.} 2006, \apjs,
  166, 470, \dodoi{10.1086/506525}

\bibitem[{{Richards} {et~al.}(2011){Richards}, {Kruczek}, {Gallagher}, {Hall},
  {Hewett}, {Leighly}, {Deo}, {Kratzer}, \& {Shen}}]{Richards11}
{Richards}, G.~T., {Kruczek}, N.~E., {Gallagher}, S.~C., {et~al.} 2011, \aj,
  141, 167, \dodoi{10.1088/0004-6256/141/5/167}

\bibitem[{{Rodriguez-Gomez} {et~al.}(2015){Rodriguez-Gomez}, {Genel},
  {Vogelsberger}, {Sijacki}, {Pillepich}, {Sales}, {Torrey}, {Snyder},
  {Nelson}, {Springel}, {Ma}, \& {Hernquist}}]{RodriguezGomez15}
{Rodriguez-Gomez}, V., {Genel}, S., {Vogelsberger}, M., {et~al.} 2015, \mnras,
  449, 49, \dodoi{10.1093/mnras/stv264}

\bibitem[{{Rubin} {et~al.}(2018){Rubin}, {O'Meara}, {Cooksey}, {Matuszewski},
  {Rizzi}, {Doppmann}, {Kwok}, {Martin}, {Moore}, {Morrissey}, \&
  {Neill}}]{Rubin18}
{Rubin}, K. H.~R., {O'Meara}, J.~M., {Cooksey}, K.~L., {et~al.} 2018, \apj,
  859, 146, \dodoi{10.3847/1538-4357/aaaeb7}

\bibitem[{{Rubinur} {et~al.}(2019){Rubinur}, {Das}, \& {Kharb}}]{Rubinur19}
{Rubinur}, K., {Das}, M., \& {Kharb}, P. 2019, \mnras, 484, 4933,
  \dodoi{10.1093/mnras/stz334}

\bibitem[{{Sanders} {et~al.}(1988){Sanders}, {Soifer}, {Elias}, {Madore},
  {Matthews}, {Neugebauer}, \& {Scoville}}]{Sanders88}
{Sanders}, D.~B., {Soifer}, B.~T., {Elias}, J.~H., {et~al.} 1988, \apj, 325,
  74, \dodoi{10.1086/165983}

\bibitem[{{Shen} \& {Liu}(2012)}]{Shen12}
{Shen}, Y., \& {Liu}, X. 2012, \apj, 753, 125,
  \dodoi{10.1088/0004-637X/753/2/125}

\bibitem[{{Shen} {et~al.}(2011){Shen}, {Richards}, {Strauss}, {Hall},
  {Schneider}, {Snedden}, {Bizyaev}, {Brewington}, {Malanushenko},
  {Malanushenko}, {Oravetz}, {Pan}, \& {Simmons}}]{Shen11}
{Shen}, Y., {Richards}, G.~T., {Strauss}, M.~A., {et~al.} 2011, \apjs, 194, 45,
  \dodoi{10.1088/0067-0049/194/2/45}

\bibitem[{{Shen} {et~al.}(2021){Shen}, {Chen}, {Hwang}, {Liu}, {Zakamska},
  {Oguri}, {Li}, {Lazio}, \& {Breiding}}]{Shen21}
{Shen}, Y., {Chen}, Y.-C., {Hwang}, H.-C., {et~al.} 2021, Nature Astronomy, 5,
  569, \dodoi{10.1038/s41550-021-01323-1}

\bibitem[{{Shen} {et~al.}(2023){Shen}, {Hwang}, {Oguri}, {Chen}, {Di Matteo},
  {Ni}, {Bird}, {Zakamska}, {Liu}, {Chen}, \& {Kratter}}]{Shen23}
{Shen}, Y., {Hwang}, H.-C., {Oguri}, M., {et~al.} 2023, \apj, 943, 38,
  \dodoi{10.3847/1538-4357/aca662}

\bibitem[{{Shields} {et~al.}(2012){Shields}, {Rosario}, {Junkkarinen},
  {Chapman}, {Bonning}, \& {Chiba}}]{Shields12}
{Shields}, G.~A., {Rosario}, D.~J., {Junkkarinen}, V., {et~al.} 2012, \apj,
  744, 151, \dodoi{10.1088/0004-637X/744/2/151}

\bibitem[{{Spingola} {et~al.}(2019){Spingola}, {McKean}, {Lee}, {Deller}, \&
  {Moldon}}]{Spingola19}
{Spingola}, C., {McKean}, J.~P., {Lee}, M., {Deller}, A., \& {Moldon}, J. 2019,
  \mnras, 483, 2125, \dodoi{10.1093/mnras/sty3189}

\bibitem[{{Steinborn} {et~al.}(2016){Steinborn}, {Dolag}, {Comerford},
  {Hirschmann}, {Remus}, \& {Teklu}}]{Steinborn16}
{Steinborn}, L.~K., {Dolag}, K., {Comerford}, J.~M., {et~al.} 2016, \mnras,
  458, 1013, \dodoi{10.1093/mnras/stw316}

\bibitem[{{Stocke} {et~al.}(1992){Stocke}, {Morris}, {Weymann}, \&
  {Foltz}}]{Stocke92}
{Stocke}, J.~T., {Morris}, S.~L., {Weymann}, R.~J., \& {Foltz}, C.~B. 1992,
  \apj, 396, 487, \dodoi{10.1086/171735}

\bibitem[{{Telfer} {et~al.}(2002){Telfer}, {Zheng}, {Kriss}, \&
  {Davidsen}}]{Telfer02}
{Telfer}, R.~C., {Zheng}, W., {Kriss}, G.~A., \& {Davidsen}, A.~F. 2002, \apj,
  565, 773

\bibitem[{{Urrutia} {et~al.}(2009){Urrutia}, {Becker}, {White}, {Glikman},
  {Lacy}, {Hodge}, \& {Gregg}}]{Urrutia09}
{Urrutia}, T., {Becker}, R.~H., {White}, R.~L., {et~al.} 2009, \apj, 698, 1095,
  \dodoi{10.1088/0004-637X/698/2/1095}

\bibitem[{{Urrutia} {et~al.}(2008){Urrutia}, {Lacy}, \& {Becker}}]{Urrutia08}
{Urrutia}, T., {Lacy}, M., \& {Becker}, R.~H. 2008, \apj, 674, 80,
  \dodoi{10.1086/523959}

\bibitem[{{Urrutia} {et~al.}(2012){Urrutia}, {Lacy}, {Spoon}, {Glikman},
  {Petric}, \& {Schulz}}]{Urrutia12}
{Urrutia}, T., {Lacy}, M., {Spoon}, H., {et~al.} 2012, \apj, 757, 125,
  \dodoi{10.1088/0004-637X/757/2/125}

\bibitem[{{Van Wassenhove} {et~al.}(2012){Van Wassenhove}, {Volonteri},
  {Mayer}, {Dotti}, {Bellovary}, \& {Callegari}}]{vanWassenhove12}
{Van Wassenhove}, S., {Volonteri}, M., {Mayer}, L., {et~al.} 2012, \apjl, 748,
  L7, \dodoi{10.1088/2041-8205/748/1/L7}

\bibitem[{{Vestergaard} \& {Wilkes}(2001)}]{Vestergaard01}
{Vestergaard}, M., \& {Wilkes}, B.~J. 2001, \apjs, 134, 1,
  \dodoi{10.1086/320357}

\bibitem[{{Wilson} {et~al.}(2004){Wilson}, {Henderson}, {Herter}, {Matthews},
  {Skrutskie}, {Adams}, {Moon}, {Smith}, {Gautier}, {Ressler}, {Soifer}, {Lin},
  {Howard}, {LaMarr}, {Stolberg}, \& {Zink}}]{Wilson04}
{Wilson}, J.~C., {Henderson}, C.~P., {Herter}, T.~L., {et~al.} 2004, in
  \procspie, Vol. 5492, Ground-based Instrumentation for Astronomy, ed.
  A.~F.~M. {Moorwood} \& M.~{Iye}, 1295--1305, \dodoi{10.1117/12.550925}

\bibitem[{{Yue} {et~al.}(2021){Yue}, {Fan}, {Yang}, \& {Wang}}]{Yue21}
{Yue}, M., {Fan}, X., {Yang}, J., \& {Wang}, F. 2021, \apjl, 921, L27,
  \dodoi{10.3847/2041-8213/ac31a9}

\end{thebibliography}
\bibliographystyle{aasjournal}

\end{document}